\documentclass{ws-ijmpcs}

\begin{document}

\markboth{C. Sigismondi}
{Gerbert of Aurillac astronomer and pope}

%
\catchline{}{}{}{}{}
%
\title{GERBERT OF AURILLAC: ASTRONOMY AND GEOMETRY IN TENTH CENTURY EUROPE}

\author{COSTANTINO SIGISMONDI}

\address{Sapienza University of Rome, Physics Dept., and Galileo Ferraris Institute\\
P.le Aldo Moro 5 Roma, 00185, Italy. e-mail: sigismondi@icra.it\\
University of Nice-Sophia Antipolis - Dept. Fizeau (France);\\
IRSOL, Istituto Ricerche Solari di Locarno (Switzerland)\\
}

\maketitle

\begin{history}
\received{6 Feb 2012}
\revised{Day Month Year}
\end{history}

\begin{abstract}

Gerbert of Aurillac was the most prominent personality of the tenth century: astronomer, organ builder and music theoretician, mathematician, philosopher, and finally pope with the name of Silvester II (999-1003). Gerbert introduced firstly the arabic numbers in Europe, invented an abacus for speeding the calculations and found a rational approximation for the equilateral triangle area, in the letter to Adelbold here discussed. Gerbert described a semi-sphere to Constantine of Fleury with built-in sighting tubes, used for astronomical observations. The procedure to identify the star nearest to the North celestial pole is very accurate and still in use in the XII century, when {\it Computatrix} was the name of {\it Polaris}. 
For didactical purposes the Polaris would have been precise enough and much less time consuming, but here Gerbert was clearly aligning a precise equatorial mount for a fixed instrument for accurate daytime observations. Through the sighting tubes it was possible to detect equinoxes and solstices by observing the Sun in the corresponding days. The horalogium of Magdeburg was probably a big and fixed-mount nocturlabe, always pointing the star near the celestial pole.
 
\end{abstract}

    
\ccode{PACS numbers:  01.65.+g, 43.10.Mq.}

\section{Introduction}	
Usually for the pseudo-epigraphy phenomenon, several publications are associated to a single personality of the past.
Aristotle's {\it Problemata} are not written by the Stagirite, and the same happened with the many authors of the apocryph accounts of the Gospel.
For Gerbert of Aurillac, after an enthousiastic start on the studies on his scientific production in XIX century (A. Olleris,\cite{olleris} N. Bubnov\cite{bubnov}), there has been a cooling down during the last decades of XX century, with doubts rising around his authorship on Astrolabe's works (U. Lindgren,\cite{lindgren} E. Poulle\cite{poulle,poulle2}).
Only in this XXI century, with the exception of the treatise on the measure of organ pipes already recognized as authentic in 1970 by K. J. Sachs,\cite{sachs,sigiABOB} the originality of the astronomical and mathematical works of Gerbert has been evidenced (U. Lindgren,\cite{lindgren2} M. Zuccato,\cite{zuccato} C. Sigismondi,\cite{sigiG} P. Rossi,\cite{rossi} F. G. Nuvolone\cite{nuvolone}).
The studies on Gerbert have been conducted in interdisciplinary contexts,
and some congresses are periodically celebrated in Rome, Bobbio (Italy) and in Paris, with public lectures also in Seoul,\footnote{$http://www.koreaherald.com/national/Detail.jsp?newsMLId=20090608000068$}   Rio de Janeiro\footnote{$http://www.mast.br/museudeideias\_2011.htm$}  and Beijing.\footnote{$http://ntsrvg9-2.icra.it/meetings/gx3\_sessions.htm$  click on History of Astronomy session}

\section{Gerbert's Geometry: Reducing Errors on Equilateral Triangle Area from $33\%$ to $1\%$}	

The contribution of Gerbert to mathematics is remarkable: he invented an abacus for doing fast calculations: the abacists in the first centuries of II millenium were also called {\it Gibertists}.\cite{falcolini}  
He was the first to introduce arabic numbers in Europe.\cite{nuvolone}

The "arithmetical formula" $A=1/2a \times(a + 1)$ for the area $A$ of an equilateral triangle of sides $a$ was stated by Boethius, and Gerbert proved experimentally that this formula was not correct in the letter to Adelbold.\cite{miller} 
This document merit the epithet "the first mathematical paper of the Middle Ages which deserves this name".\cite{hankel}
The previous formula gave $A=28$, while he found $A=21$ by using a triangle of side 7 and height 6. The true value is $A\sim 21.2176$.
The height of an equilateral triangle of side 7 is $h\sim 6.062$. Therefore also the final result on the area is accurate within 1 part over 100: $\Delta A/A\le 1.1\%$.
Since $\sqrt 3$, the ratio between the height of the triangle and half side is irrational, its approximation with rational numbers can be as good as wanted, but since "exempla have to be done with small numbers" Gerbert chosen the smaller numbers.  
Other choices with three integer numbers (n, m, 2n) such as $n^2+m^2=4n^2+d$ are (15,26,30) with d=1 and (41,71,82) (with d=2) for wich  $\sqrt 3 = 1.73205...$
is approximated respectively by $26/15= 1.7\bar{3}$ and $71/41=1.\overline{73107}$. 
In the letter to Adelbold, Gerbert corrected the height 26 of the "equilateral" triangle of side 30, to 25+5/7.
This correction makes this triangle similar to the one of side 7 and height 6 already seen, and does not improve the approximation of $\sqrt 3$ obtained with 26/15, and because of that this letter of Gerbert has been criticized and considered as an example of the low level of mathematical accuracy achieved in the tenth century.\cite{miller2} 
Gerbert considered the triangle of side 7 and heigth 6 as an approximation good enough to the equilateral one.
From the example of side 30 and height 26, reduced by Gerbert to 25+5/7, it is possible to argue that he could not compute the $\sqrt 3$ and approximate it, otherwise he would have chosen 26/15 a value closer than 1 part over 1000, instead of 6/3.5=(25+5/7)/15. 
Conversely the choice of using a rule concerning smaller numbers (i.e. 6 and 7) may be interpreded as of practical sense:
the approximation achieved with 6 and 7 is only $\sim 1\%$ below the correct value, and it is worthless to use the following rational approximations of 26 and 15 or 71 and 41. According to this interpretation Gerbert had to know the Pythagoras' theorem,\cite{rossi} and his approach would be more similar to the approach of modern physics, which is basically numerical.
For an accuracy of $1\%$ in the calculation of the area of an equilateral triangle the triangle with height 6 and sides 7 is good enough, while the arithmetical rule, based only on the side's length, would imply an unaccetable error of $+33\%$, and its elimination could have been the main goal of Gerbert in writing this letter.
The vibrant school of geometry in Li\`{e}ge at the beginning of XI century (Franco of Li\`{e}ge found iterative algorithms to calculate square roots of numbers) existed as a direct result of Gerbert's teaching.\cite{nancy}


\begin{figure}
\centerline{\includegraphics[width=1\textwidth,clip=]{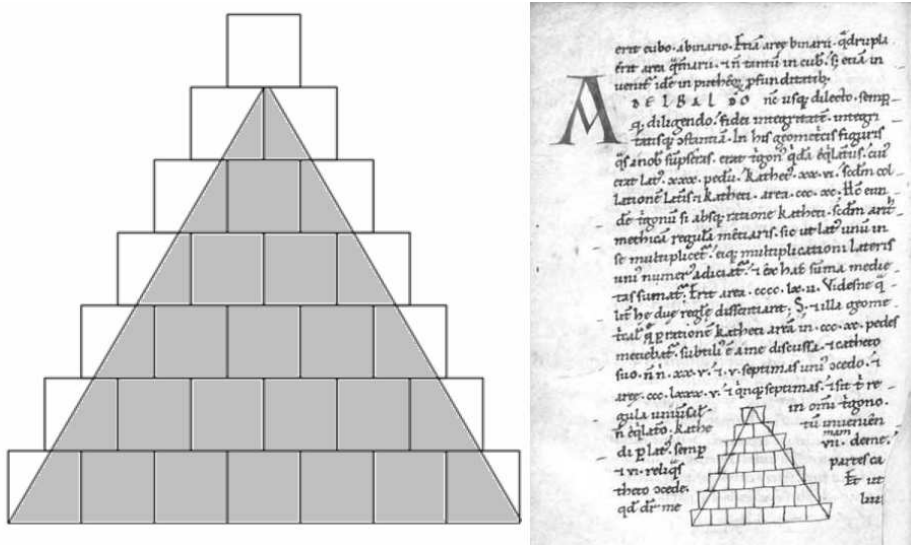}}
\caption{The approximation of the equilateral triangle: left according to Bubnov 1899, right in a manuscript of XII century. The uneven squares' dimensions induced probably the copist to put the vertex on the 7th floor of this pyramid, instead of the 6th as in the exemplum of Gerbert in the letter to Adelbold. {\it To comprehend this more clearly lend your eyes to this figure, and always remember me} wrote Gerbert to him.}
\label{Fig. 1}
\end{figure}

\begin{figure}
\centerline{\includegraphics[width=1\textwidth,clip=]{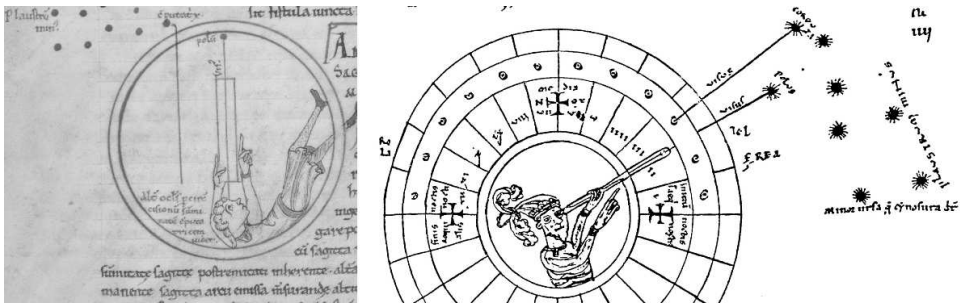}}
\caption{Left: manuscript of Avranches BN 235, f. 32, XII century. The use of a {\it fistula} to point the pole's star called {\it Polus} and the other eye aims at another star called {\it Computatrix}, which is our modern {\it Polaris}. The constellation represented is the Ursa Minor, also called {\it Cynosura} or the {\it Plaustrum minus} (the {\it Little Dipper}). Right: reproduction of the manuscript 173 of Chartres, destroyed in a II World War bombing, showing the same map, with unambiguos indications of the two stars {\it Polus} and {\it Computatrix}.}
\label{Fig. 2}
\end{figure}

\begin{figure}
\centerline{\includegraphics[width=1\textwidth,clip=]{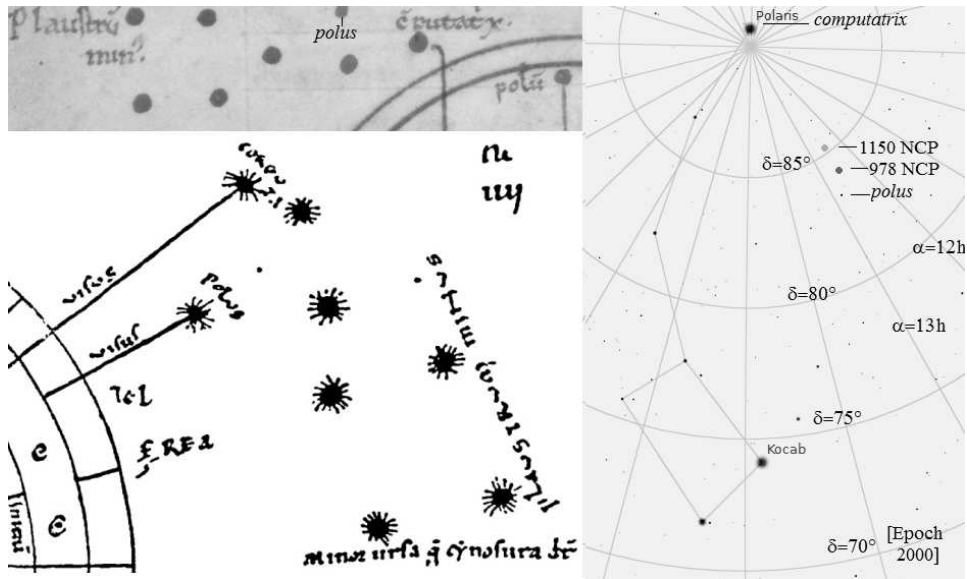}}
\caption{Enlarged details of figure 2, to show the correspondence between the star {\it Polus} and the star HR4893. In both manuscripts the location of the {\it Polus} is near the center of the arch of three stars concluding with the {\it Polaris}, i.e. the {\it Computatrix}. On the star map there are the locations of the North Celestial Pole (NCP) in the years 978 and 1150, respctively the years of the letter to Constantine about the semi-sphere, and of the two manuscripts of Avranches and Chartres. The star HR4893 was only $1^o$ apart from NCP, obtaining the same pointing precision of modern mechanical mounts for amateurs (e.g. VIXEN Super-Polaris {\rm http://www.company7.com/vixen/history.html}).}
\label{Fig. 3}
\end{figure}



\section{Gerbert Astronomer: an Equatorial Fixed Mount for Solar Observations and the Pole's Star}	

Gerbert describes a semi-sphere to Constantine of Fleury. This sphere has
sighting tubes, used for astronomical observations. The procedure
to identify the star nearest to the North Celestial Pole is very
accurate. The star, HR 4893, of magnitude $m_v=5.3$ is visible to the naked eye under dark skies.\cite{sigiG}
The description made by Gerbert helps to find it in case of loosing the star map,
because there were a few stars of this luminosity around the North celestial pole.
For didactical purposes the {\it Polaris} would have been
precise enough and much less time consuming, but here Gerbert is
clearly aligning a precise equatorial mount which had to be logically
set up as a fixed instrument for accurate daytime observations.
Through the sighting tubes it was possible to detect equinoxes and
solstices by observing the Sun in the corresponding days. 


The bishop Thietmar of Merseburg (975-1018) reported in his {\it Chronicon} 
for the year 1014 about an "horalogium" left by Gerbert in Magdeburg: {\it "In Magdeburg horalogium fecit, recte illud constituens secundum quandam stellam, nautarum ducem, quam consideravit per fistulam miro modo".} 
and still visible at his times.

This permanent istrument could have been the one described to Constantine of Fleury:\cite{sigiG} or a nocturlabe\cite{poulle} fixed and big enough to be visible to all visitors of Magdeburg, as reported by Thietmar.\cite{thietmar} 

The {\it Polaris}, probably after Gerbert's instructions to point the {\it Polus}, was called {\it Computatrix} during XII century, as attested in the manuscripts of figure 2.
H. Michel\cite{michel} supposed that {\it Computatrix} was our {\it Polaris} and {\it Polus} the star HR 4893 [in the Harvard Revised stellar catalogue or HD 112028 (in the Henry Draper catalogue, also known as 32H Camelopardalis, from the Hevelius' catalogue], that I also have seen with naked eyes\cite{sigiNC} in Camelopardalis. The same hypothesis was made by Wiesenbach.\cite{Wiese1,Wiese2}
The star {\it Polus}, is not recorded in Ptolemy's Almagest because at his time it was not on the celestial pole, but it is visible to the naked eye, and it is represented as smaller (fainter) than the others and in the correct location in this schematic map.
Once the pole is centered by the {\it fistula} only another star is necessary to compute the hour of the night in a given day of the year. 
Note that the line drawn between {\it Polus} and {\it Computatrix-Polaris} contained also the North Celestial Pole of 978 and 1150. This makes the nocturlabe\cite{oestmann} used by Gerbert very accurate to determine the hour of the night.

\end{document}